\documentclass{article}

\usepackage{arxiv}

\usepackage[utf8]{inputenc} % allow utf-8 input
\usepackage[T1]{fontenc}    % use 8-bit T1 fonts
\usepackage{hyperref}       % hyperlinks
\usepackage{url}            % simple URL typesetting
\usepackage{booktabs}       % professional-quality tables
\usepackage{amsfonts}       % blackboard math symbols
\usepackage{nicefrac}       % compact symbols for 1/2, etc.
\usepackage{microtype}      % microtypography
\usepackage{lipsum}
\usepackage{amsmath}
\usepackage{graphicx}
\usepackage{adjustbox}
\usepackage{multirow}
\usepackage{multicol}
\usepackage{xcolor}
\usepackage{authblk}
\graphicspath{ {./} }

\title{Mediastinal lymph nodes segmentation using 3D convolutional neural network ensembles and anatomical priors guiding}

\author[1,2]{David Bouget}
\author[1]{Andr{\'e} Pedersen}
\author[1]{Johanna Vanel}
\author[3]{Haakon O. Leira}
\author[1]{Thomas Lang{\o}}
\affil[1]{Department of Medical Technology\\
  SINTEF\\
  Trondheim, Norway}
\affil[2]{Center for Innovative Ultrasound Solutions, NTNU, Trondheim, Norway}
\affil[3]{ Department of Thoracic Medicine\\
  St. Olavs hospital\\
  Trondheim, Norway }

\begin{document}
\maketitle
\begin{abstract}
\textbf{Purpose:} As lung cancer evolves, the presence of enlarged and potentially malignant lymph nodes must be assessed to properly estimate disease progression and select the best treatment strategy. Following the clinical guidelines, estimation of short-axis diameter and mediastinum station are paramount for correct diagnosis. A method for accurate and automatic segmentation is hence decisive for quantitatively describing lymph nodes.\\
\textbf{Methods:} In this study, the use of 3D convolutional neural networks, either through slab-wise schemes or the leveraging of downsampled entire volumes, is investigated. Furthermore, the potential impact from simple ensemble strategies is considered. As lymph nodes have similar attenuation values to nearby anatomical structures, we suggest using the knowledge of other organs as prior information to guide the segmentation task.\\
\textbf{Results:} To assess the segmentation and instance detection performances, a 5-fold cross-validation strategy was followed over a dataset of $120$ contrast-enhanced CT volumes. For the $1\,178$ lymph nodes with a short-axis diameter $\geq10$\,mm, our best performing approach reached a patient-wise recall of $92$\%, a false positive per patient ratio of 5, and a segmentation overlap of $80.5$\%. The method performs similarly well across all stations.\\
\textbf{Conclusion:} Fusing a slab-wise and a full volume approach within an ensemble scheme generated the best performances. The anatomical priors guiding strategy is promising, yet a larger set than four organs appears needed to generate an optimal benefit. A larger dataset is also mandatory, given the wide range of expressions a lymph node can exhibit (i.e., shape, location, and attenuation), and contrast uptake variations.
\end{abstract}

% keywords can be removed
\keywords{3D segmentation, Ensemble, Attention U-Net, Mediastinal lymph nodes, CT}

\section{Introduction}
\label{intro}
Lung cancer manifests itself through the development of malignant tumors characterized by uncontrolled cell growth in tissues of the lung. As cancer evolves, its growth can spread beyond the lung and reach nearby anatomical structures such as lymph nodes, causing them to grow in size~\cite{falk2010}. Lymph nodes are routinely assessed by clinicians to monitor disease progression, establish cancer diagnosis, or simply evaluate the effect of therapeutics given their propensity to enlarge under the effect of many pathologies. As defined by the Response Evaluation Criteria In Solid Tumors (RECIST) guidelines~\cite{eisenhauer2009new}, a lymph node with a short-axis diameter of at least $10$\,mm in an axial CT slice is likely to be malignant and represent a clinical interest~\cite{schwartz2009evaluation}.
Accurate clinical or pretreatment stage classification of lung cancer leads to optimal patient outcomes and improved prognostication. Regional lymph node maps are mandatory to facilitate consistent and reproducible lymph node designations, and are regularly issued by the International Association for the Study of Lung Cancer (IASLC)~\cite{el2014international}. The regional map defines fourteen different general anatomic locations, also called stations, precisely described by a set of guidelines articulated around neighbouring anatomical structures in the mediastinum (e.g., blood vessels, airways, or ligaments). In practice, a lymph node is assigned its station by an expert radiologist, according to its relative position with respect to nearby anatomical structures.
In lung cancer diagnosis, a chest contrast-enhanced Computed Tomography (CT) scan is most frequently favored, and represents the gold-standard modality. In absence of distant metastasis, enlarged lymph nodes identified on chest CT demand a verification procedure, either through endobronchial ultrasound (EBUS)~\cite{sorger2017multimodal} or mediastinoscopy, to ascertain the severity and aggressiveness of the cancer. However, the manual segmentation of lymph nodes in the mediastinal area is tedious, highly time-consuming, and requires trained experts. The process is inherently subject to intra/inter-observer variability depending on the time allocated to perform the task, level of concentration, and quality of the CT scan~\cite{mcerlean2013intra}. In addition, challenges arise from the relatively similar attenuation between lymph nodes and surrounding structures (e.g., esophagus, azygos vein, or other vessels), especially impactful when poor contrast enhancement is exhibited. Last but not least, lymph nodes manifest themselves through extensive variations in shape, size, texture, and location.
An automatic method is therefore of high importance for facilitating the tasks of lymph node segmentation and standardized measurements computation (e.g., short-axis diameter, station) in order to assist the clinical team in making the best cancer staging.

\setlength{\parindent}{5ex} Advances in machine learning, and more specifically deep learning, has boosted the performance in image segmentation, and fully convolutional neural networks~\cite{long2015fully} have been widely accepted for medical image segmentation~\cite{ronneberger2015u,zhou2017fixed}. While deep learning-based methods thrive on everyday-life data, access to sufficiently large and annotated training datasets in the medical field represents a known bottleneck. In recent years, medical image segmentation, from CT volumes amongst other modalities, has known tremendous progress, driven by subsequent challenges where annotated datasets were made publicly available to foster research and challenge medical communities. For instance, the segmentation of moderately large organs (e.g., liver, pancreas, or kidneys) from abdominal CT volumes was initially investigated in multiple challenges~\cite{kavur2020chaos,simpson2019large}. Such organs present the advantage of being easy to delineate, without the required assistance of an expert radiologist, to create trustworthy datasets. Initially, most current deep learning architectures were designed to operate using a 2D input, hence requiring CT volumes to be segmented one sectional image (i.e., slice) at a time~\cite{vesal20192d,wang2019abdominal}. Yet in practice, a radiologist would scroll through the CT volume and across all views to properly identify the full extent of a given anatomical structure. Naturally, anatomical borders can be only discernible from subtle change in texture or shape, not always visible in consecutive 2D slices.
To overcome such limitations, attempts have been made to analyze all three views at the same time in a 2.5D fashion~\cite{zhuang2016multi}, or by using multiple 2D patches around the segmentation target~\cite{setio2016pulmonary}. Fortunately, with the increasing capacity of GPUs, and the ability for neural network architectures to seamlessly adapt to different dimensions, studies over 3D medical images emerged~\cite{cciccek20163d}. Attempts to leverage full resolution CT volumes are still held back by their huge memory footprint. The rising bottlenecks comprise a high model complexity (i.e., number of parameters), longer training times, and overfitting issues. Nevertheless, access to local 3D context can be gained by processing the raw volume in a slab-wise or block-wise fashion~\cite{kamnitsas2016deepmedic}.
For the joint tasks of segmentation and instance detection, popular approaches have been widely used over 2D images~\cite{he2017mask,redmon2018yolov3}. While such architectures are powerful and could translate well to the 3D domain, setting up the region proposal layer is too memory-expensive and training time would be a challenge.

In the literature, mediastinal lymph nodes have predominantly been studied for detection purposes and on occasion for segmentation, most often leveraging only a contrast-enhanced CT volume. In their initial work, Oda et al. employed standard machine learning technique to perform lymph node detection whereby hand-crafted features were extracted using a Hessian-based strategy~\cite{oda2017hessian}. In a follow-up study, a two-step detection algorithm was proposed, based on an intensity targeted radial structure tensor and blob-like structure enhancement filters~\cite{oda2017automated}. Similarly, Paing et al. ~\cite{paing2019automatic} proposed a pipeline mixing traditional image features extraction (i.e., threshold, watershed, and hessian eigenvalues) fed to a classification 3D neural network separating lymph node candidates from false lesions. In mediastinal and abdominal contexts, Nogues et al. focused on the segmentation of lymph node clusters~\cite{nogues2016automatic}. The cope with reduced intensity and texture contrast amongst collocated lymph nodes, 2D holistically-nested neural networks were proposed as a solution to perform embedded edge detection. Structured optimization techniques were then promoted to refine the produced imprecise segmentation, such as conditional random fields and graph cuts. Roth et al.~\cite{roth2014new} presented a two-stage pipeline to perform detection of all the potentially malignant lymph nodes. All voxels were first classified as to either belonging to the lymph node class or the background by use of blobness and circular transforms. Fed by the results of the first stage, a convolutional neural network, trained in a 2.5D fashion, produced the final set of lymph node candidates.
Overall, the proposed methods were partly or fully built upon ad-hoc components or required strongly hand-crafted features, which represents a clear limitation given the impossibility to assume fixed intensity thresholds from input volumes with varying quality. In addition, prior assumptions over the shape of a lymph node were made (i.e., roundish blob) and are often incorrect. In consecutive works, Liu et al. addressed the topic of mediastinal lymph node detection and station mapping using the previously described two-stage pipeline~\cite{liu2014mediastinal,liu2016mediastinal}. Eight anatomical structures were additionally segmented to help in the station mapping task. Each lymph node candidate was assigned a station based on its centroid location w.r.t. surrounding structures and following the IASLC guidelines. Five stages and more than half an hour were necessary to produce the final segmentation and instance detection results. The use of hand-crafted features and multiple steps also limit the method's ability to generalize and deployment in practice. Fully end-to-end approaches appears more adequate to address both limitations at once. To that end, Oda et al.~\cite{oda2018dense} proposed a method to perform lymph node detection and segmentation using a 3D U-Net. To improve their performances and reduce data imbalance issues, they also proposed to include four anatomical structures (i.e., lungs, airways, aortic arch and branches, and pulmonary arteries). While a high true positive detection rate was reached, the reported false positive per patient rate of 17 appeared quite high. Including up to fourteen anatomical structures, in addition to the lymph nodes, Bouget et al. proposed a pipeline operating in 2D~\cite{bouget2019semantic}. The suggested approach combined the pixel-wise segmentation capability of a U-Net and improved instance detection capability from Mask R-CNN. Validated only on a dataset of $15$ patients, a recall of $75$\% was reached for $9$ false positives per patient on average, but the lack of global information was detrimental to obtain competitive results.
Facing the issue of limited access to ground truth, Li et al.~\cite{li2020deep} devised a weakly supervised method generating bounding boxes and pixel-wise pseudo-mask from the RECIST annotations. A U-Net architecture was used to produce the initial pixel-wise segmentation and a deep reinforcement-based component was coupled for optimization. State-of-the-art performance was attained over thoracoabdominal lymph nodes, reaching a Dice score of $77$\%.
Many previous studies performed their experiments using the only open-source dataset, first introduced by Roth et al.~\cite{roth2014new}. The dataset contains $90$ CT volumes featuring mediastinal lymph nodes, together with manual annotations provided for most lymph nodes with a short-axis diameter larger than $10$\,mm. However, the annotations are often sparse, and every smaller but visible lymph node was left unsegmented, making it suboptimal to use straight off the shelf for fully supervised pixel-wise segmentation purposes.

In another line of work, a few studies have focused on leveraging both contrast-enhanced CT and PET/CT modalities concurrently. Only the malignant lymph nodes are exhibiting uptake in the PET volume leaving the benign ones unnoticeable. For simple pixel-wise semantic segmentation, Xu et al. proposed to use the DeepLabv3+ architecture to benefit from atrous spatial pyramid pooling operating at various grid scales, improving boundary segmentation~\cite{xu2020focal}. Focal loss was further studied to let the network focus on the difficulty-to-segment voxels and prevent overfitting on the other category of nonchallenging voxels. Zhu et al.~\cite{zhu2020lymph} proposed a multi-branch detection-by-segmentation network to perform segmentation and detection of malignant lymph nodes. An effective distance-based gating approach was developed in their proposed framework, replicating protocols conducted by oncologists in daily practice. Unfortunately, a PET/CT scan is usually acquired at a later stage during the diagnostic process, after cancer suspicion is raised from the analysis of the contrast-enhanced CT scan. As such, an optimal processing of this initial CT volume is mandatory before considering the PET/CT modality.

In this study, our contribution are the following: (i) the investigation of slab-wise, full volume, and ensemble strategies for semantic segmentation in 3D to benefit from local and global context, (ii) validation studies showing segmentation and detection performances with respect to lymph nodes' short-axis diameter and station providing insights on the impact of training data variability, (iii) the largest annotated mediastinal lymph nodes dataset with $120$ patients and close to $3\,000$ lymph nodes, partially available in open-access.

\section{Data}
\label{sec:dataset}
\begin{figure}[ht]
\centering
\includegraphics[scale=0.79]{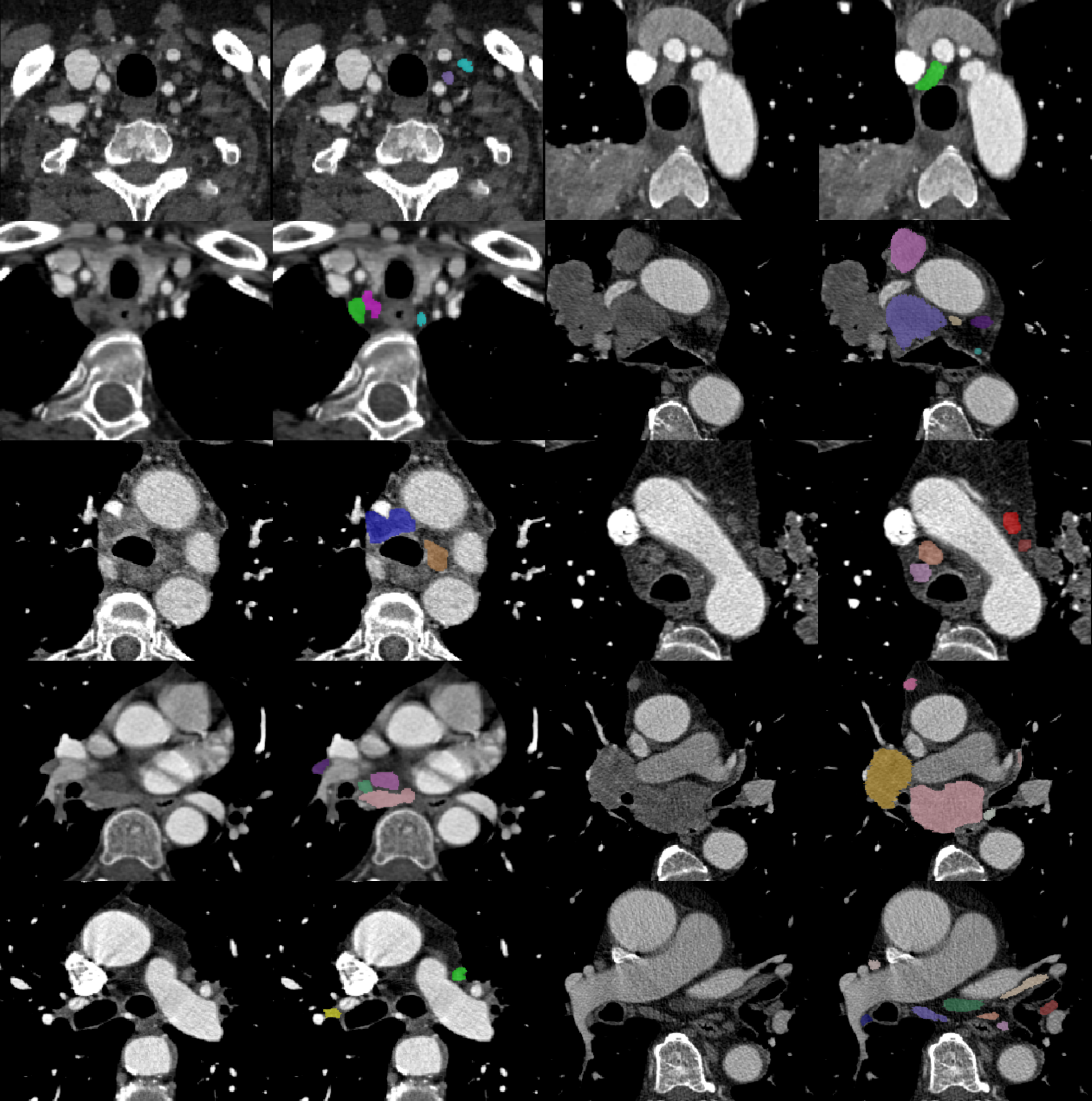}
\caption{Illustrations of the manual annotations over the dataset. For each pair, the left image represents the raw CT clipped to $[-150;500]$\,HU, and the right image represents the same CT slice with the manually annotated lymph nodes overlaid in color. From top to bottom, and left to right, each example features lymph nodes in different stations.}
\label{fig:dataset-illu}
\end{figure}

In this work, a dataset of $120$ contrast-enhanced CT volumes was assembled, featuring lung cancer patients exhibiting malignant lymph nodes and coming from two different sources. First, the $89$ patients with mediastinal lymph nodes from the NIH open-source dataset~\cite{roth2014new} were gathered, which are publicly available for download~\footnote{https://wiki.cancerimagingarchive.net/display/Public/CT+Lymph+Nodes}. The remaining $31$ volumes were acquired on lung cancer patients at the Department of Thoracic Medicine, St. Olavs hospital, Trondheim University Hospital, Norway.

Overall, CT volume dimensions are covering $[487; 512]\times[441; 512]\times[56; 854]$\,voxels, and the voxel size ranges $[0.58; 0.97]\times[0.58; 0.97]\times[0.5; 5.0]$\,mm$^{3}$. An average CT volume is $[511\times511\times628]$ pixels with a spacing of $[0.79\times0.79\times0.99]$\,mm$^3$.
For the CT volumes acquired at the St. Olavs hospital, lymph nodes' manual annotations were performed by an expert thoracic radiologist. Regarding the CT volumes from the NIH dataset, the available annotations were used as a starting point and manually refined by a medical trainee under the supervision of the expert. Following the RECIST criterion, malignancy for a lymph node is considered for a short-axis diameter larger than $10$\,mm. Using the \textit{regionprops} method from the Scikit-Image Python package, the short-axis diameter was computed for each annotated lymph node. A total of $2\,912$ lymph nodes are featured in our dataset, with $1\,178$ having a short-axis diameter larger than $10$\,mm, $767$ having a short-axis diameter in the range $[7, 10[$\,mm, and $967$ with a short-axis diameter smaller than $7$\,mm. A set of annotated lymph nodes from our dataset is illustrated in Fig~\ref{fig:dataset-illu}.

\begin{figure}[t]
\centering
\includegraphics[scale=0.55]{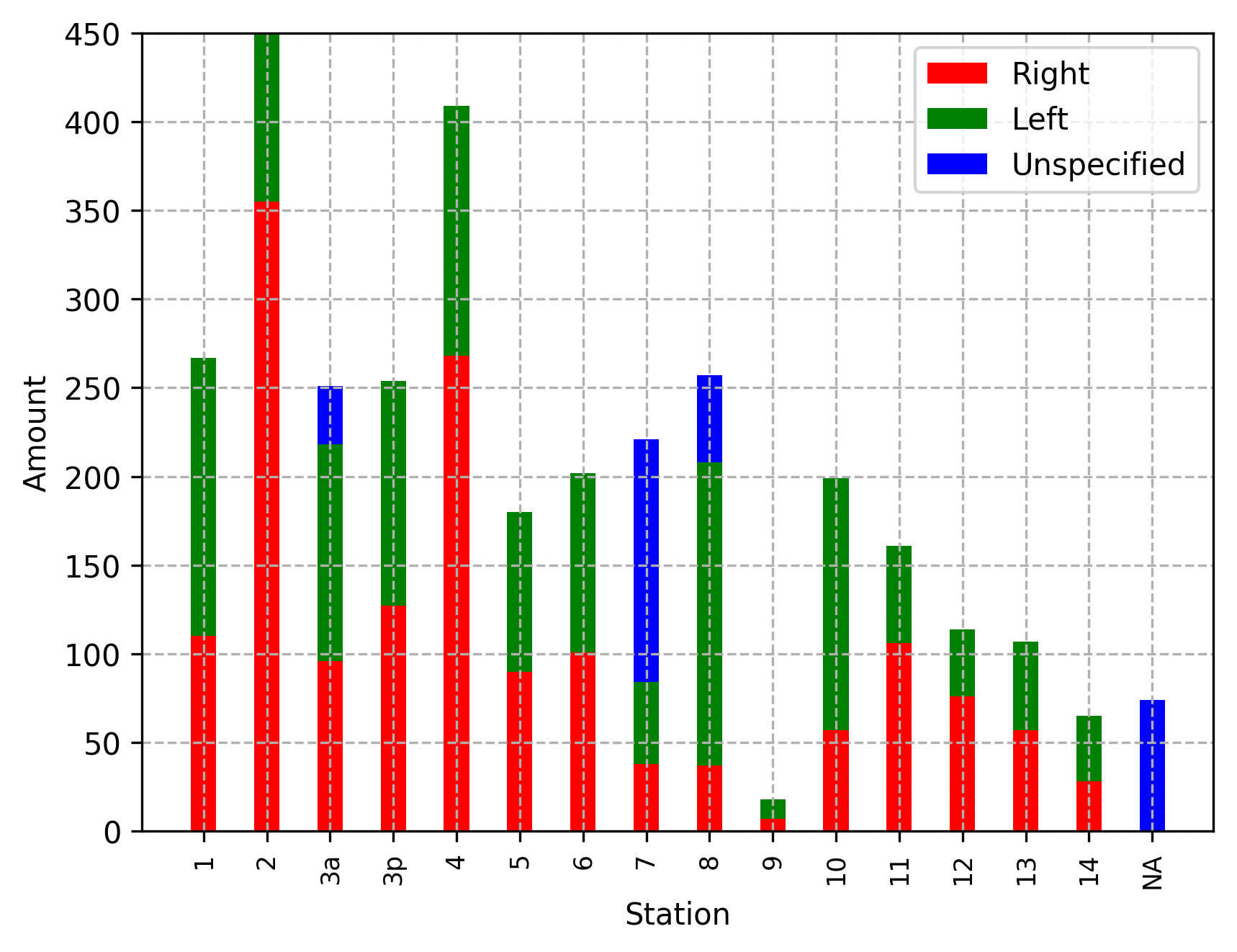}~\includegraphics[scale=0.55]{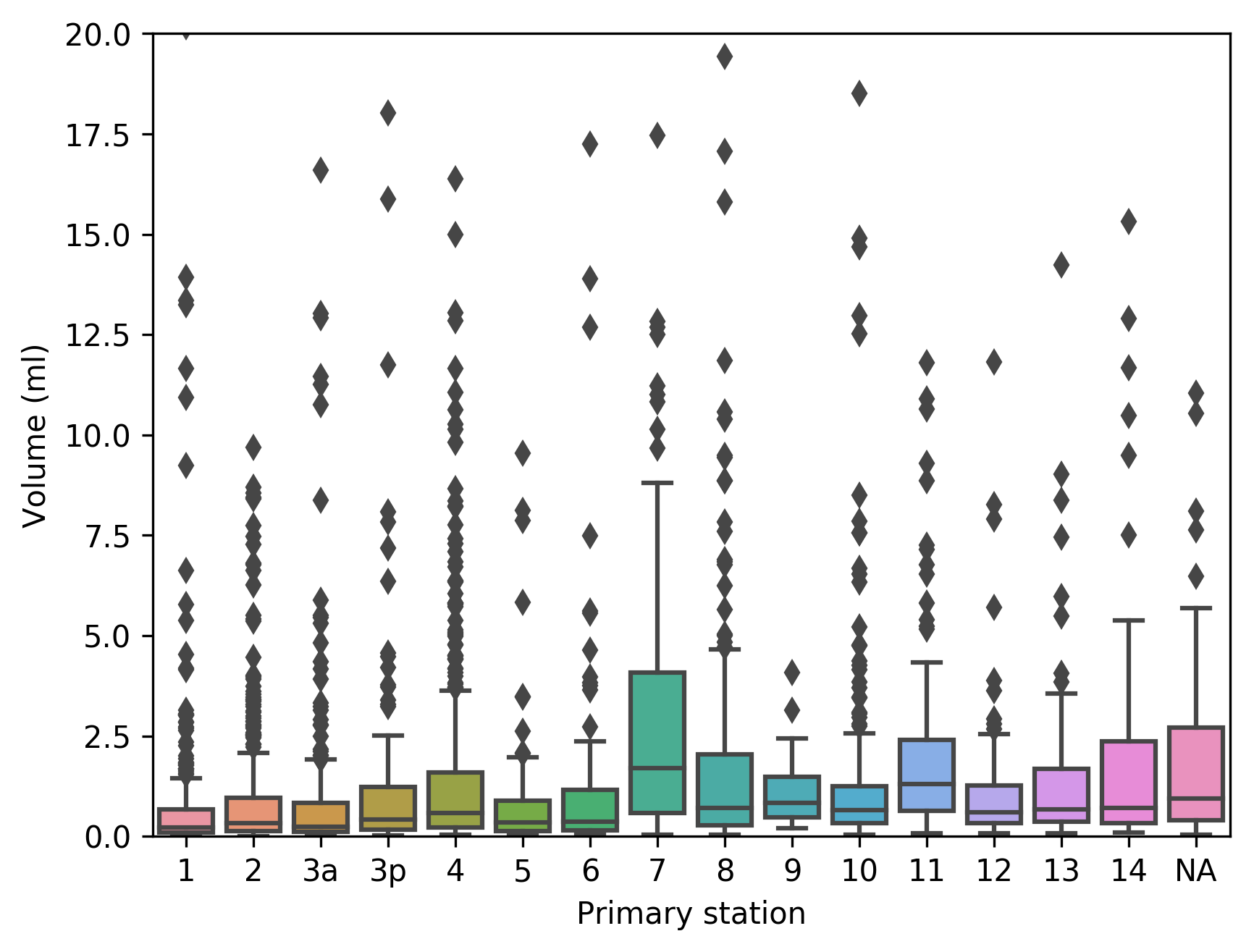}
\caption{Dataset statistics regarding the primary station distribution (to the left), and the volume distribution within each primary station (to the right). Each station is uniquely represented and the lateralization distribution is not directly shown. Lymph nodes not mapped to any IASLC station are regrouped under the NA category.}
\label{fig:dataset-ln-stations}
\end{figure}

Following the IASLC guidelines, each lymph node was assigned its station by a medical trainee under supervision of the expert. The exhaustive list of stations' number and name is as follows: 1) Low cervical, supraclavicular, and sternal notch nodes, 2) Upper paratracheal, 3a) Prevascular, 3p) Retrotracheal, 4) Lower paratracheal, 5) Subaortic, 6) Para-aortic, 7) Subcarinal, 8) Paraesophageal below carina, 9) Pulmonary ligament, 10) Hilar, 11) Interlobar, 12) Lobar, 13) Segmental, and 14) Sub-segmental. Most stations can also be subdivided between left and right side, but in the rest of the paper we will refer to a station by its main number. Given the large variability in lymph nodes' expression (e.g., shape and size), overlaps across multiple stations are common and multiple station assignments were required. Occasionally, lymph nodes could not be directly mapped to the guidelines or were outside the mediastinum scope, and were thus left station-less (e.g., around the heart, below the base of the lungs, or at the base of the neck). Overall, $74$ lymph nodes were left unassigned, $1\,256$ were overlapping at least two stations and $379$ were overlapping at least three stations. In terms of volume, lymph nodes are ranging $[0.01, 234.72]$\,ml with a mean value of $1.98\pm6.81$\,ml. Lymph nodes statistics with respect to volume and primary station are illustrated in Fig.~\ref{fig:dataset-ln-stations}. The primary station distribution is represented to the left, with unassigned lymph node grouped in the NA category. A large imbalance can be noticed, especially between station 2 and 9, and it can be noted that on average lymph nodes are more present on the right side. In some cases, the decision could not be made regarding lateralization, represented as unspecified (in blue). Volume-wise, the distribution across primary stations is relatively homogeneous but the plot cannot be interpreted unbiasedly as often more than one station is overlapped by a lymph node, especially for the larger ones.

\section{Methods}
\begin{figure}[!t]
\centering
\includegraphics[scale=0.5]{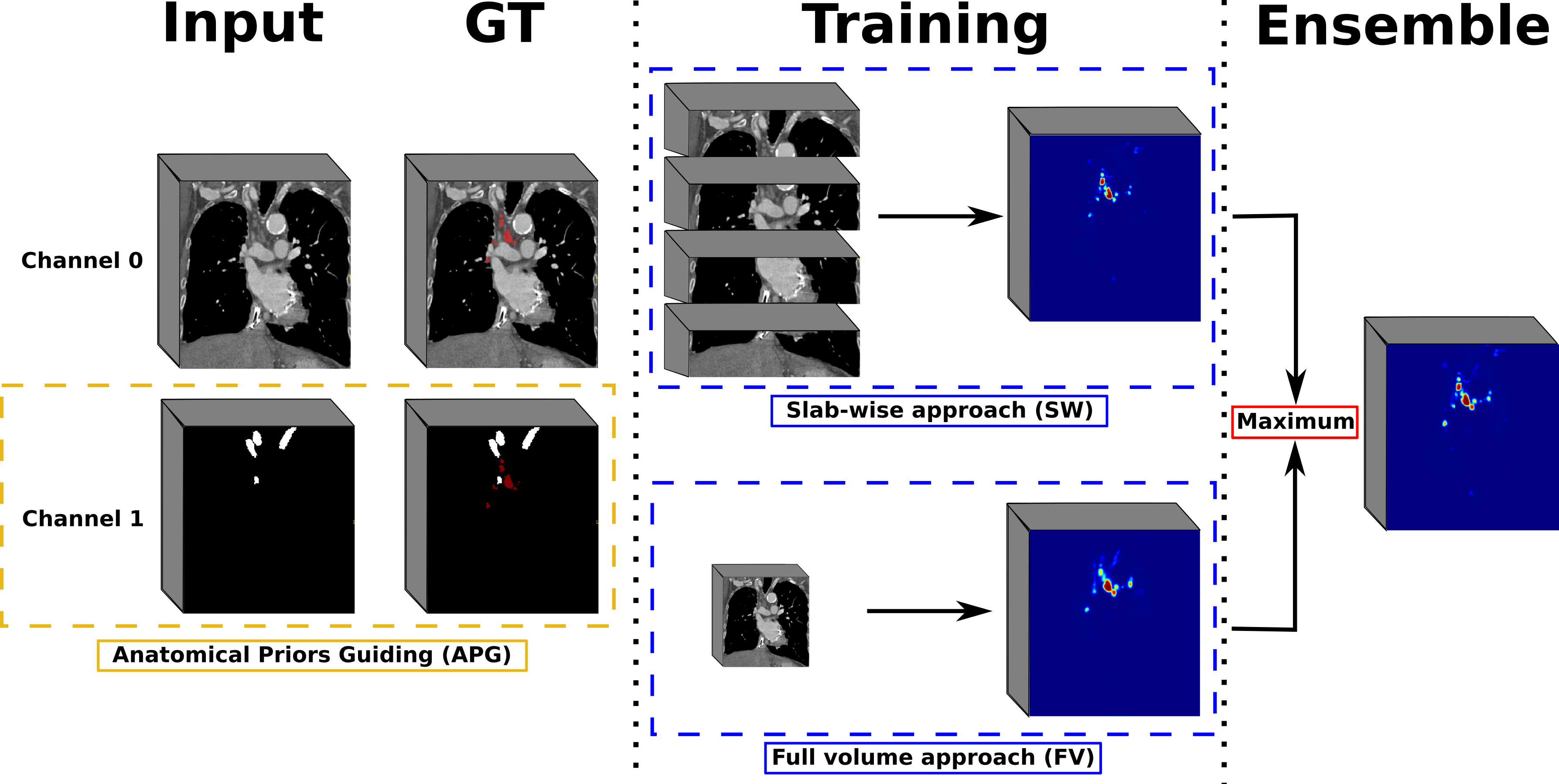}
\caption{Overall representation of the proposed approach, where a slab-wise and a full volume approach are fused in an ensemble fashion, with the possibility to include anatomical priors.}
\label{fig:overall-achitecture}
\end{figure}

First, a description of the neural network architectures and designs used in this study is provided in Section~\ref{subsec:architectures}. Then, the possibility to leverage anatomical priors to guide the network during training is introduced, together with our proposed ensemble strategy, in Section~\ref{subsec:anatomical-prior}. Finally, the different preprocessing steps and selected training strategies are detailed in Section~\ref{subsec:train-strats}. An overview of our proposed approach is given in Fig.~\ref{fig:overall-achitecture}.

\subsection{Architectures and designs}
\label{subsec:architectures}
For properly segmenting and detecting instances of potentially collocated lymph nodes, the most promising alternatives described in the literature are multi-task architectures the likes of Mask R-CNN~\cite{he2017mask} or YOLACT~\cite{bolya2019yolact}. Even though  extremely good performance was documented from studies carried out in the 2D domain, fewer studies were applied to the 3D domain as the transition presents many limitations ranging from memory overload, to complexity to generate enough sensible proposals.
When dealing with 3D medical volumes, too sizable to fit on GPU memory, two main lines of work can be identified: (i) splitting the volume in a slab-wise manner (SW) and (ii) using the full volume at a lower resolution (FV). In the slab-wise strategy, the objective is to benefit locally from a high-resolution while generating contextual features on some extent of global information, which is correlated to the slab size. In the full volume strategy, overall contextual features can be computed to model all spatial relationships between each visible anatomical structure. However, due to memory limitation and time constraint for training, keeping the initial resolution is not feasible yet and degrading the spatial resolution is needed as opposed to the slab-wise strategy.
Over the previous years, ample focus has been dedicated to improving 2D/3D pixel-wise segmentation performances over well-established backbone architecture such as ResNet~\cite{he2016deep} and U-Net~\cite{ronneberger2015u}, providing near radiologist-level performances on some medical image analysis tasks~\cite{bai2017human,liao2019evaluate}. The common pitfall for such encoder-decoder architectures is some extent of details' loss provoked by strided convolutions and pooling operations along the encoding path to progressively enlarge the field-of-view. The constant challenge lies in optimally using global and contextual information from high-level features and border information from low-level features to resolve small details~\cite{sang2020pcanet}.
To compensate for the loss of details in the encoding path, previous studies have non-exhaustively covered multi-scale investigations employing the input volume at down-sampled scales in each encoder block~\cite{abraham2019novel}, enlarging the receptive fields using atrous convolutions and pyramid spatial pooling~\cite{chen2017rethinking}, or by crafting multi-scale feature maps encoding jointly low-level and high-level semantic information in a powerful representation~\cite{sinha2020multi}. Similarly, instead of computing the loss simply from the last decoder step intermediate feature maps generated at each level of the architecture can be leveraged. Feature maps from hidden layers of a deep network can serve as a proxy to improve the overall segmentation quality and sensitivity of the model, while alleviating the problem of vanishing gradients~\cite{lee2015deeply}.
In a different line of work, attention mechanisms are able to identify salient image regions and amplify their influence while filtering away irrelevant information from other regions, making the prediction more contextualised~\cite{jetley2018learn}. Attention is optimally coupled to each level during the decoding path and can be designed to capture features' dependencies spatially, channel-wise, or across any other dimension~\cite{fu2019dual}. Interestingly, all those suggested concepts can be seamlessly integrated into the current CNN architectures considered as backbones and enable complete end-to-end training.

\label{sec:methods}

In this work, given the inherent 3D nature of the various anatomical structures in the mediastinal area, we chose to focus on improving the lymph nodes' pixel-wise segmentation using 3D U-Net~\cite{cciccek20163d} as backbone architecture (denoted as UNet). Our UNet design has been set to seven levels and $[8, 16, 32, 64, 128, 256, 256]$ as filter sizes for each level respectively. For studying the impact of concepts like deep supervision or attention, we chose the Attention-Gated U-Net (AGUNet) and Dual-Attention Guided U-Net (DAGUNet) architectures as previously introduced~\cite{bouget2021meningioma}. For those, the design was set to five levels and $[32, 64, 128, 256, 512]$ as filter sizes for each level respectively.

\subsection{Anatomical knowledge priors and model ensembles}
\label{subsec:anatomical-prior}
In the mediastinal area, many anatomical structures have attenuation values in the same range as the lymph nodes (e.g., esophagus, azygos vein) often without clear in-between boundaries. To assist the model in better differentiating between lymph nodes and surrounding similar-looking structures, anatomical knowledge priors can be injected to serve as prior knowledge during training. To that end, each training sample can be built as a combination of the raw CT positioned in channel 0, and a binary mask containing anatomical priors placed in channel 1. With this approach, the no-go zones defined in the anatomical priors' mask should be identified to prevent the model generating high probabilities, hence decreasing the false positive ratio.

Multiple models operating on different input shapes or focusing on different aspects during training can be ensembled to generate a better consensus and hence improve the final prediction map~\cite{feng2020brain}. Global context and local refinement can virtually be obtained separately at the cost of longer training and inference time, and higher model complexity.
Regarding the task of lymph nodes' segmentation, the two main strategies considered (i.e., slab-wise and full volume) operate on different domains and present each inherent strengths and weaknesses. Since training a model using the whole CT at a high resolution is not feasible, performing ensembling appears to be a competitive alternative solution. In this study, we opted for a straightforward approach whereby a maximum operator is applied pixel-wise over the probability maps resulting from a model trained following each of the two main strategies. By doing so,
the newly created probability map is biased toward a better recall at the detriment of the precision since more voxels will have a higher probability as belonging to the lymph node class.

\subsection{Training strategies}
\label{subsec:train-strats}
To prepare the training samples, all CT volumes were preprocessed identically using the following series of steps: (i) resampling to an isotropic spacing of $1$\,mm$^{3}$ using spline interpolation order 1 from NiBabel~\footnote{https://github.com/nipy/nibabel}, (ii) lung-cropping using a pre-trained network~\cite{hofmanninger2020automatic} in order to generate the tightest bounding box around the mediastinal area, (iii) resizing to the network's input resolution using spline interpolation order 1, and (iv) intensity clipping to the range $[-250, 500]$\,HU followed by normalizing to the range $[0, 1]$.

For the slab-wise investigations performed in this study, we chose to wield slabs of 32 slices and 64 slices along the z-axis denoted as SW32 and SW64 respectively, for an axial resolution of $256\times192$\,pixels. In order to generate the collection of training samples, a stride parameter of 8 was employed whereby two consecutive slabs would share 24 slices and 56 slices respectively. The value was empirically chosen as a good trade-off by which models can be trained in a decent amount of time, generalize well, and circumvent overfitting hurdles.
Regarding the full volume approach, a drastic down-sampling was imperative given the GPU memory limitations, therefore a new resolution of $128\times128\times144$\,voxels was chosen.
Investigations performed using anatomical priors guiding were performed solely with the full volume approach, hence with a final training sample dimension of $128\times128\times144\times2$\,voxels. The four following structures were used as part of the anatomical priors: esophagus, azygos vein, subclavian arteries, and brachiocephalic veins.

For the data augmentation strategy, the following transforms were applied to each input sample with a probability of $50$\%: horizontal and vertical flipping, random rotation in the range $[-20^{\circ}, 20^{\circ}]$, translation up to 10\% of the axis dimension, zoom between $[80, 120]\%$ in the axial plane.
Given the large variability in lymph nodes' shape and location, and the relatively limited total number of patients, the training samples were randomly assigned to their fold at a patient-level without any other consideration. All models were trained from scratch using the Adam optimizer with an initial learning rate of $10^{-3}$, and training was stopped after $30$ consecutive epochs without validation loss improvement. The main loss function used was the class-average Dice, excluding the background, and batches of size 8 were used for slab-wise training, while batches of size 1 with 32 accumulated gradient steps were used for full volume training. The concept of accumulated gradient enables training with larger batch sizes, whereby N batches are run sequentially using the same model weights for calculating the gradients. When the N steps are performed, the model weights are properly updated using the accumulated sum of gradients.

\section{Validation studies}
\label{sec:validation}
In this work, we aim at assessing the lymph nodes' pixel-wise segmentation and partial instance detection performances, leaving aside the task of disambiguation between collocated instances. The dataset was split randomly into 5 folds at the patient level and a 5-fold cross-validation performed whereby three folds were used for training, one for validation, and one for testing. As described in Section~\ref{subsec:gt-preproc}, the initial ground truth was slightly refined to adjust for the inability to distinguish between collocated lymph nodes. The selected measurements and metrics used are then introduced in Section~\ref{subsec:meas-met} while the validation studies are detailed in Section~\ref{subsec:val-studies}.

\subsection{Ground truth preprocessing}
\label{subsec:gt-preproc}
\begin{figure}[t]
\centering
\includegraphics[scale=0.94]{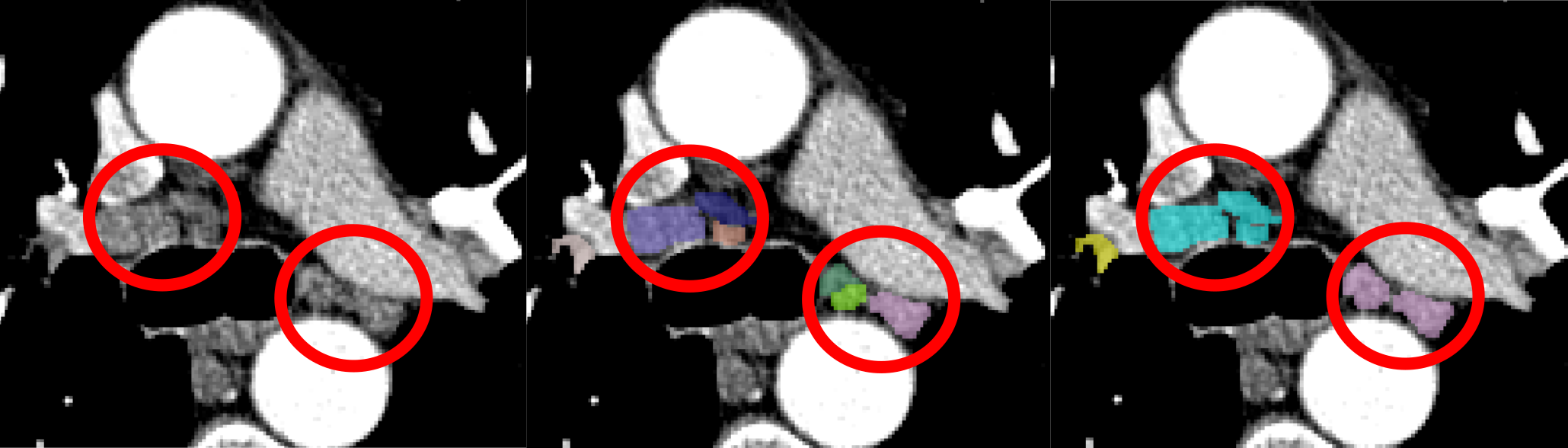}
\caption{Illustration of the preprocessing performed over the ground truth annotation to prepare for instance detection performance assessment. The raw CT slice is shown to the left, the original ground truth with 7 lymph nodes is visible in the middle, and the final ground truth with 3 lymph nodes is presented to the right. The two transformed lymph node clusters are circled in red.}
\label{fig:cc-gt-illu}
\end{figure}

In order to assess instance detection performances (e.g., recall, precision) over results from a pure pixel-wise segmentation neural network, some adjustments must be performed.
From the resulting probability map, and after thresholding using the optimal value, a connected components approach is operated in order to generate lymph node instances. However, collocated lymph nodes would only produce one instance using such an approach. In addition, the same behaviour is to be expected for close-by lymph nodes, given some noisiness in the final binary mask.
To reflect the inability of our method to optimally perform instance segmentation and reduce the bias in reported performances, we applied the same connected components method on the original ground truth. Each new lymph node cluster was assigned with all primary stations for each individual element composing the cluster. An illustration is provided in Fig.~\ref{fig:cc-gt-illu}, where two clusters of collocated lymph nodes (circled in red) are visible.

\subsection{Measurements and metrics}
\label{subsec:meas-met}
To assess the segmentation performance, the Dice score is computed between the ground truth and a binary representation of the probability map generated by a trained model. The binary representation is computed for ten different equally-spaced probability thresholds (PT), in the range $[0, 1]$.
After identification of the best PT for each model, the instance detection performance is assessed in turn. A connected components approach, used over the resulting binary map, coupled to a pairing strategy is executed to identify matches between a ground truth and a detection mask, based on the Dice score computation.

The Dice score, reported in \%, is used to assess the quality of the pixel-wise segmentation at a patient level. To refine and highlight the reported segmentation performance at a lymph node level, the Dice score over true positive instances is further computed (noted Dice-TP). Finally, discrepancies are expected between paired lymph node instances from the use of a connected components approach over noisy masks. To prevent the results from being heavily penalized by such shortcomings, we propose to report in complement the total percentage of the ground truth to have been correctly segmented at a lymph node level (noted GT-perc).

To report instance detection performance, the recall as a global metric across all patients is supplied, as well as a per-patient recall (noted Recall-PW) to take into account the high variability in the amount of lymph nodes from patient to patient. Regarding precision, the number of false positives per patient (FPPP) is mentioned.

Lastly, the pure inference speed and the total elapsed time required to generate predictions for a new patient are reported (in s), obtained with both CPU or GPU support. The operations required to prepare the data to be sent through the network, to initialize the environment, to load the trained model, and to reconstruct the probability map in the referential space of the original volume are accounted for. The experiment has been repeated five consecutive times over the same CT volume, using a representative high-resolution case of $512\times512\times767$\,voxels with $0.68\times0.68\times0.5$\,mm$^{3}$ spacing.
All measurements are reported with mean and standard deviation.

\subsection{Studies}
\label{subsec:val-studies}
Experiments using various combinations of architectures and training strategies (as introduced in Section~\ref{sec:methods}) were carried out, and the name given to each experiment is a descriptive concatenation. Architectures to choose from are: regular U-Net (UNet), attention-gated U-Net (AGUNet), and dual attention guided U-Net (DAGUNet). Training strategies to choose from are: slab-wise with 32 slices (SW32), slab-wise with 64 slices (SW64), full volume (FV), and using anatomical priors guiding (APG). Experiments using ensembles are specifically mentioned.

\textit{Slab-wise performance analysis:} On average, twelve days were necessary to train one model in a slab-wise fashion. Under such circumstances, comparing all slab-wise methods using a 5-fold cross-validation approach was realistically unachievable. The objective of this first study is to show limited performance variability across the five different folds using the UNet-SW32 approach. All models trained in a slab-wise fashion are then compared over the first fold only to identify the best performing approach.

\textit{Overall performance comparison:} Averaged over the 5 folds, a comparative analysis is performed between the different methods considered and across the whole set of proposed metrics. In this study, all lymph nodes featured in our dataset are considered without restrictions over the short-axis diameter or primary station. A total of $2\,122$ lymph nodes or lymph node clusters, after ground truth preprocessing, are taken into account.

\textit{Performance analysis over lymph nodes' characteristics:} For the best performing model identified in the previous validation study, an in-depth analysis is carried out with respect to the short-axis diameter and primary station.
Three divisions are used to highlight the performance with regard to the short-axis diameter, around the thresholds of $7$\,mm and $10$\,mm.
Given our choice of preprocessing steps to generate the training samples, many lymph nodes featured in station 1 (i.e., above or at the top of the lungs) were never seen by the network and hence could not be segmented. As such, two main groups are considered: (i) all stations even for the NA category (noted all stations), and (ii) all stations except 1 and NA (noted relevant stations).

\textit{Ground truth quality assessment over a benchmark subset:} Out of the $31$ CT volumes acquired at the St. Olavs hospital and featured in our dataset, $15$ have been part of previous studies and are publicly available~\cite{reynisson2015airway}. To assess the quality of the ground truth, a second expert with a background as a thorax radiologist was asked to look over the segmentation and assigned stations for this subset. To judge the inter-rater variability regarding station assignment, and given that lymph nodes can overlap multiple stations, we defined three grades: (i) perfect if both annotators were in agreement regarding the primary station, (ii) good if annotators agreed on the multiple stations overlapped but disagreed on the primary station, and (iii) bad if the second expert assigned a different and unrelated primary station.\\
In addition to the ground truth made publicly available, segmentation and instance detection performances obtained over the $15$ patients are separately reported for benchmark purposes.

\section{Results}
\label{sec:results}

Models were trained using an HP desktop: Intel Xeon @3.70 GHz, 62.5 GiB of RAM, NVIDIA Quadro P5000 (16GB), and a regular hard drive. Implementation was done in Python 3.6 using \texttt{TensorFlow} v1.13.1, Cuda 10.0, and the Imgaug Python library for the data augmentation methods~\cite{imgaug}.
Trained models, inference code, and ground truth files are available at~\url{https://github.com/dbouget/ct_mediastinal_structures_segmentation}.

\subsection{Slab-wise performance analysis}
For each of the five folds, performances obtained using the UNet-SW32 method are summarized in Table~\ref{tab:results-cross-fold-variability}. The recall and Dice score values achieved over the first fold are above the average performances by 3\%. Performances across each fold are relatively stable and contained within 4\% of the average values. The highest magnitude in difference exists for the patient-wise recall with 12\% between folds 3 and 4. However, it is worth noting that fold 4 exhibits the smallest amount of single or collocated lymph nodes, around 25\% fewer than in any other fold. Comparing the different slab-wise training schemes based on the first fold only appears trustworthy enough to identify the best performing approach in a reasonable amount of time.

\begin{table}[!ht]
\centering
\caption{Segmentation and instance detection performances collected for the five folds with the UNet-SW32 approach. The \# LN corresponds to the number of lymph nodes or clusters in each fold.}
\adjustbox{max width=\textwidth}{
\begin{tabular}{c|c|c|c|c|c|c|c|c|c}
Fold & PT & \# LN & Dice & Recall & Recall-PW & FPPP & GT-Perc \tabularnewline
\hline
1 & 0.3 & 459 & $67.34\pm14.49$ & 56.43 & $58.16\pm16.49$ & $4.04\pm1.85$ & $65.58\pm9.45$ \tabularnewline
2 & 0.3 & 410 & $64.47\pm11.25$ & 55.12 & $57.63\pm16.94$ & $5.83\pm3.34$ & $71.40\pm10.7$ \tabularnewline
3 & 0.3 & 451 & $61.88\pm14.83$ & 47.01 & $48.85\pm17.73$ & $5.83\pm2.94$ & $66.56\pm12.84$ \tabularnewline
4 & 0.3 & 330 & $63.21\pm16.40$ & 55.15 & $60.71\pm18.93$ & $9.37\pm4.56$ & $70.80\pm12.68$ \tabularnewline
5 & 0.3 & 472 & $64.58\pm16.27$ & 49.36 & $50.93\pm18.21$ & $4.08\pm2.26$ & $63.99\pm14.45$ \tabularnewline
\hline
Total & 0.3 & 2122 & $64.27\pm14.63$ & 52.40 & $55.23\pm17.98$ & $5.85\pm3.65$ & $67.68\pm12.32$ \tabularnewline
\end{tabular}
}
\label{tab:results-cross-fold-variability}
\end{table}

\begin{table}[b]
\caption{Overall performances obtained with the different slab-wise training configurations, for the first fold only. The fourth experiment, indicated by *, used a lower input resolution of $192\times128$\,voxels.}
\adjustbox{max width=\textwidth}{
\begin{tabular}{l|c|c|c|c|c|c|c|c|c|c}
Experiment & PT & Dice & Dice-TP & GT-Perc & Recall & Recall-PW & FPPP \tabularnewline
\hline
(i) UNet-SW32 & 0.3 & \boldmath{$67.34\pm14.49$} & \boldmath{$53.49\pm14.21$} & $65.58\pm09.45$ & \boldmath{$56.43$} & \boldmath{$58.16\pm16.49$} & $4.04\pm1.85$ \tabularnewline
(ii) UNet-SW64 & 0.4  & $65.86\pm14.18$ & $52.46\pm11.01$ & $62.77\pm08.23$ & $52.72$ & $56.94\pm15.66$ & $4.57\pm2.04$ \tabularnewline
(iii) AGUNet-SW64 & 0.2 & $64.24\pm13.55$ & $53.37\pm10.83$ & $77.12\pm09.51$ & $53.38$ & $55.49\pm15.34$ & \boldmath{$3.00\pm1.41$} \tabularnewline
(iv) AGUNet-SW64* & 0.2 & $62.86\pm13.24$ & $50.31\pm13.00$ & \boldmath{$78.55\pm09.72$} & $51.42$ & $53.96\pm16.82$ & $
3.74\pm2.16$ \tabularnewline
\end{tabular}
}
\label{tab:results-slabwise}
\end{table}

Performances over the first fold for the different models trained using a slab-wise strategy are reported in Table~\ref{tab:results-slabwise}. The best performing method regarding the Dice score and recall is the slab-wise U-Net architecture with 32 slices (experiment (i)), reaching up to 58\% patient-wise recall for all lymph nodes. Doubling the number of slices in the slab worsened all scores by about 2\%, probably indicating the need for more training samples (cf. experiments (i) and (ii)). Using more advanced architectures such as the attention-gated U-Net are showing improved precision with one less false positive per patient and improved GT-Perc, demonstrating improved pixel-wise segmentation around detected lymph nodes (cf. experiments (ii) and (iii)). However, this increase is accompanied by deteriorated recall and patient-wise recall values. By downsampling the input resolution, performances are diminished by 2\% at most as can be seen between experiments (iii) and (iv). Nonetheless, the minimal drop in performance for a faster training appears as a good trade-off to investigate training schemes or architecture designs in a more reasonable amount of time.

\subsection{Overall performance comparison}
\begin{figure}[b]
\centering
\includegraphics[scale=2.55]{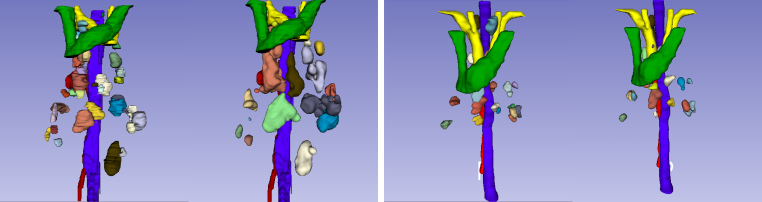}
\caption{Representation in 3D of the results for two patients where the ground truth is shown to the left and the model output to the right. The four anatomical structures used for the anatomical priors guiding are also represented: esophagus (blue), azygos vein (red), subclavian arteries (yellow), and brachiocephalic veins (green).}
\label{fig:proba-results-illu-3D}
\end{figure}

For the overall performance comparison study, results have been averaged over the five folds and are reported in Table~\ref{tab:results-seg-overall}.
The best slab-wise approach is competitive in terms of recall when compared to full volume approaches, as can be seen between experiments (i) and (ii). In addition, and given the higher spatial resolution, the slab-wise approach is able to top pixel-wise segmentation performances with up to 64\% overall Dice score compared to 60\% with the best full volume approach (cf. experiments (i) and (iv)). Using more advanced architectures, in combination with the use of anatomical priors guiding, lead to a 3\% increase in recall and patient-wise recall together with a reduction of almost 1 FPPP (cf. experiments (ii) and (iii)). When ensembling a slab-wise and a full volume approach, recall and patient-wise recall performances are similar (cf. experiments (iii) and (vi)). However, only through ensembling can all the metrics reach close to their maximum simultaneously, and as such experiment (vi) is deemed to be our best performing method. The ensemble of a slab-wise U-Net using 32 slices and a full volume attention-guided U-Net using anatomical priors reaches a patient-wise recall of $58.8\%$, a Dice score of $64\%$, an extent of segmented ground truth up to $71\%$, for 5 false positives per patient. A 3D illustration for two patients, showing the ground truth and detected lymph nodes side-by-side, is provided in Fig.~\ref{fig:proba-results-illu-3D}, and where the four organs used for the anatomical priors guiding are also featured.

\begin{table}[!ht]
\caption{Overall performance comparison obtained for the different experiments carried out, averaged across the five folds. The abbreviations are: regular U-Net (UNet), attention-gated U-Net (AGUNet), dual attention guided U-Net (DAGUNet), anatomical priors guiding (APG), full volume (FV), and slab-wise with 32 slices (SW32).}
\adjustbox{max width=\textwidth}{
\begin{tabular}{l|c|c|c|c|c|c|c|c|c|c|}
Experiment & PT & Dice & Dice-TP & GT-Perc & Recall & Recall-PW & FPPP \tabularnewline
\hline
(i) UNet-SW32 & 0.3 & $64.27\pm14.63$ & \boldmath{$53.93\pm13.79$} & $67.68\pm12.32$ & $52.40$ & $55.23\pm17.98$ & $5.85\pm3.65$ \tabularnewline
(ii) AGUNet-FV & 0.2 & $59.48\pm13.48$ & $45.21\pm16.93$ & $72.73\pm12.17$ & $52.21$ & $55.86\pm19.07$ & $4.48\pm2.98$ \tabularnewline
(iii) AGUNet-FV-APG & 0.2 & $57.64\pm13.53$ & $42.13\pm15.88$ & \boldmath{$75.07\pm12.35$} & \boldmath{$55.42$} & $58.04\pm19.95$ & $4.48\pm2.95$ \tabularnewline
(iv) DAGUNet-FV-APG & 0.4 & $60.01\pm15.78$ & $45.89\pm16.51$ & $65.81\pm14.27$ & $50.80$ & $54.21\pm19.17$ & $4.19\pm2.87$ \tabularnewline
(v) Ensemble (i) \& (ii) & 0.4 & \boldmath{$65.01\pm12.75$} & $53.59\pm14.97$ & $69.70\pm11.44$ & $52.59$ & $56.02\pm18.15$ & $5.12\pm2.96$ \tabularnewline
(vi) Ensemble (i) \& (iii) & 0.4 & $64.00\pm13.21$ & $51.89\pm15.32$ & $71.01\pm11.46$ & $55.23$ & \boldmath{$58.79\pm19.72$} & $5.23\pm2.80$ \tabularnewline
(vii) Ensemble (i) \& (iv) & 0.6 & $64.16\pm14.78$ & $50.98\pm14.75$ & $62.02\pm12.74$ & $50.47$ & $53.66\pm19.13$ & \boldmath{$4.05\pm2.27$} \tabularnewline
\end{tabular}
}
\label{tab:results-seg-overall}
\end{table}

\begin{figure}[!t]
\centering
\includegraphics[scale=0.65]{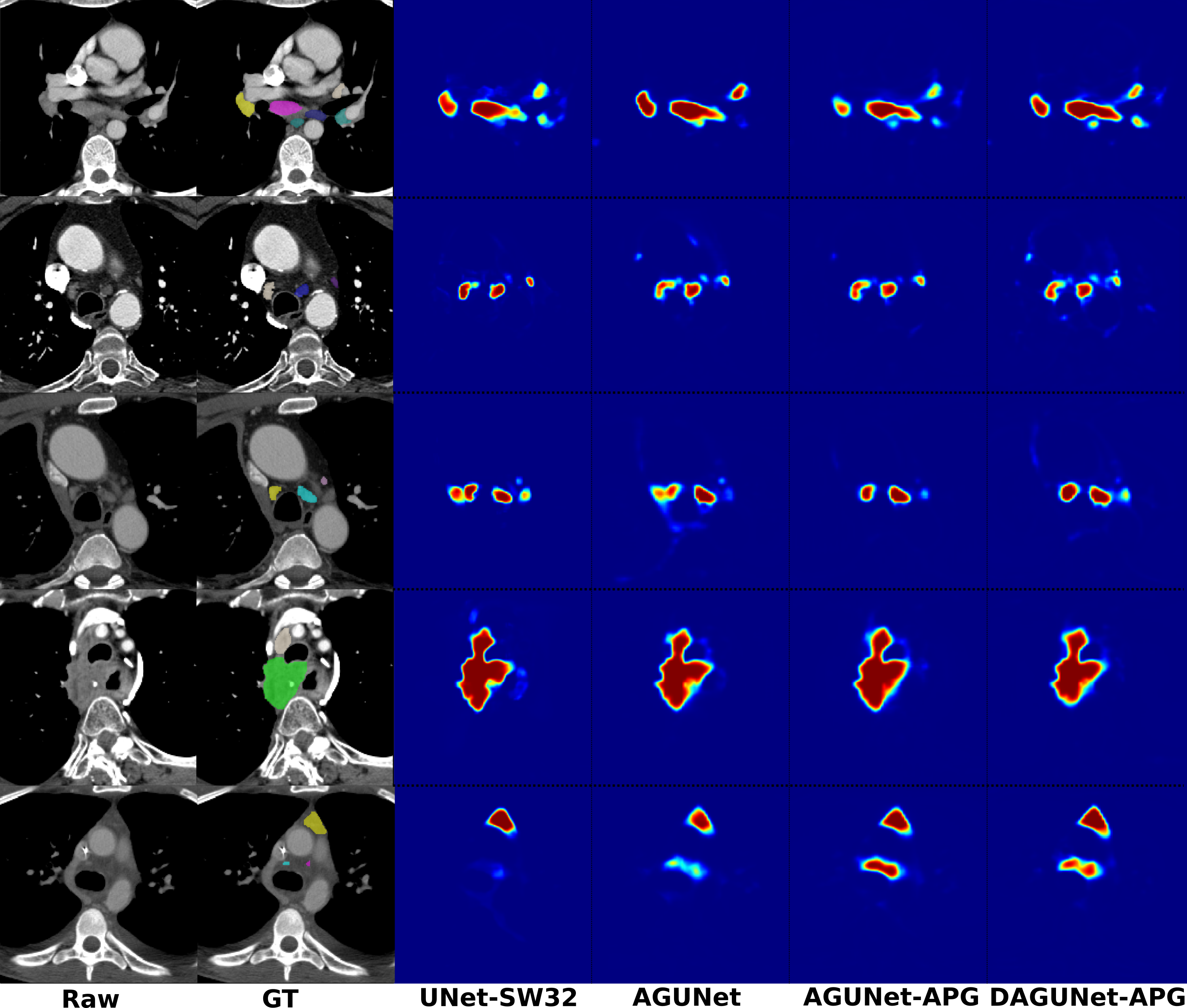}
\caption{Prediction examples for a patient from each fold, one per line, for the four main designs: UNet-SW32, AGUNet, AGUNet-FV-APG, and DAGUNet-FV-APG.
Cool colors indicate low confidence in the lymph node class and warm colors indicate high confidence.}
\label{fig:proba-results-illu}
\end{figure}

For five different patients, one being featured per row and sampled from each fold, visual comparisons are provided in Fig.~\ref{fig:proba-results-illu} between the four main designs. The probability maps are set to the range $[0, 1]$ where cool colors indicate low confidence in the lymph node class and warm colors indicate high confidence, and the CT images are shown in the range $[-150, 500]$\,HU. In the first and fourth row, the segmentation performance is similar across all methods for the displayed lymph nodes. The effect of anatomical prior guiding can be witnessed in the third row, over the azygos vein to the left of the yellow lymph node. In the first two methods without guiding the upper part of the vein is by mistake partially segmented as being a lymph node. On the opposite, the azygos vein is left totally unsegmented when using anatomical prior guiding as can be shown with the last two methods. In the last row, the contrast is slightly different than from other patients, resulting in lymph nodes more challenging to detect. While the top lymph node (colored in yellow) is similarly segmented across all methods, two other lymph nodes are featured in the ground truth (colored in blue and magenta) with varying extent of segmentation. Given the difficulty to visually identify those lymph nodes, judging whether the manual ground truth has been under-segmented or the last two methods are over-segmenting, is extremely challenging.

\begin{table}[!b]
\centering
\caption{Quantification of false positive segmentation over the four anatomical structures used for priors guiding. For each of the four main designs, Dice scores averaged across the $120$ patients are reported.}
\adjustbox{max width=\textwidth}{
\begin{tabular}{l|c|c|c|c|}
Experiment & Esophagus & Azygos vein & Subclavian arteries & Brachiocephalic veins \tabularnewline
\hline
(i) UNet-SW32 & $3.54\%$ & $3.54\%$ & $0.49\%$ & $1.27\%$ \tabularnewline
(ii) AGUNet-FV & $6.87\%$ & $7.26\%$ & $1.20\%$ & $2.57\%$ \tabularnewline
(iii) AGUNet-FV-APG & $2.03\%$ & $1.89\%$ & $0.90\%$ & $0.60\%$ \tabularnewline
(iv) DAGUNet-FV-APG & $1.61\%$ & $1.64\%$ & $0.85\%$ & $0.48\%$ \tabularnewline
\end{tabular}
}
\label{tab:apg-benefit}
\end{table}

To quantify and fully appreciate the effect of anatomical priors guiding on the pixel-wise segmentation performance rather than just visually, a supplementary analysis has been completed. The Dice scores between a model prediction and each of the four organs making the anatomical priors are reported in Table~\ref{tab:apg-benefit}. Reported scores are predominantly lower for the two experiments making use of the priors when compared with the other two experiments. Without the use of priors, the slab-wise approach seems to fair better, with twice as low scores, than the full volume approach (cf. experiments (i) and (ii)). The higher spatial resolution from the slab-wise approach being surely profitable to distinguish between similar-looking voxels belonging to different anatomical structures. The benefit is even more apparent for over the esophagus and azygos vein, with a drastic reduction in false positive segmentation (cf. experiments (ii) and (iii)). 

Regarding the total processing time required to process a new CT volume with a single model, an average of $3.5\pm0$\,minutes has been calculated, using GPU support. For ensembling, models were run sequentially without any specific optimization, leading to a total processing time of $7.3\pm0.2$\,minutes. Interestingly, a similar performance is reached using only the CPU. Most of the processing time is dedicated to the different resampling and resizing operations, given the large resolution of the input CT volume, and performed by default on the CPU. Regarding pure inference, including model loading, only $8.4\pm0.4$\,seconds are necessary on average.

\subsection{Performance analysis over lymph nodes' characteristics}

\iffalse
\begin{table}[!ht]
\caption{Segmentation performances obtained with our best performing method for the three lymph node categories based on short-axis value. The first four measurements (left-most columns) are reported for all lymph nodes while the last four measurements (right-most columns) are reported for the lymph nodes belonging to relevant IASLC stations.}
\adjustbox{max width=\textwidth}{
\begin{tabular}{c||c|c|c|c||c|c|c|c|}
 \multirow{2}{*}{Short-axis} & \multicolumn{4}{c||}{All stations} & \multicolumn{4}{c|}{Relevant stations} \tabularnewline
 \cline{2-9}
 & Dice-TP & GT-Perc & Recall & Recall-PW & Dice-TP & GT-Perc & Recall & Recall-PW \tabularnewline
\hline
\hline
All & $51.89\pm15.32$ & $71.01\pm11.46$ & 55.23 & $58.79\pm19.72$ & $51.68\pm15.54$ & $71.59\pm11.77$ & $61.50$ & $65.46\pm20.14$ \tabularnewline
$\geq7\,mm$ & $55.03\pm15.31$ & $74.20\pm12.05$ & 75.63 & $79.59\pm18.41$ & $54.80\pm15.52$ & $74.69\pm12.46$ & $80.53$ & $83.33\pm18.05$ \tabularnewline
$\geq10\,mm$ & $59.02\pm16.48$ & $80.04\pm12.60$ & 85.73 & $88.96\pm14.75$ & $58.54\pm16.99$ & $80.55\pm12.54$ & $90.73$ & $92.10\pm14.43$ \tabularnewline
\end{tabular}
}
\label{tab:results-size-stations}
\end{table}
\fi

\begin{table}[!ht]
\caption{Segmentation performances obtained with our best performing method for the three lymph node categories based on short-axis value. The first four measurements (left-most columns) are reported for all lymph nodes while the last four measurements (right-most columns) are reported for the lymph nodes belonging to relevant IASLC stations.}
\adjustbox{max width=\textwidth}{
\begin{tabular}{c|cccc|cccc|}
 \multirow{2}{*}{Short-axis} & \multicolumn{4}{c|}{All stations} & \multicolumn{4}{c|}{Relevant stations} \tabularnewline
 \cline{2-9}
 & Dice-TP & GT-Perc & Recall & Recall-PW & Dice-TP & GT-Perc & Recall & Recall-PW \tabularnewline
\hline
All & $51.89\pm15.32$ & $71.01\pm11.46$ & 55.23 & $58.79\pm19.72$ & $51.68\pm15.54$ & $71.59\pm11.77$ & $61.50$ & $65.46\pm20.14$ \tabularnewline
$\geq7\,mm$ & $55.03\pm15.31$ & $74.20\pm12.05$ & 75.63 & $79.59\pm18.41$ & $54.80\pm15.52$ & $74.69\pm12.46$ & $80.53$ & $83.33\pm18.05$ \tabularnewline
$\geq10\,mm$ & $59.02\pm16.48$ & $80.04\pm12.60$ & 85.73 & $88.96\pm14.75$ & $58.54\pm16.99$ & $80.55\pm12.54$ & $90.73$ & $92.10\pm14.43$ \tabularnewline
\end{tabular}
}
\label{tab:results-size-stations}
\end{table}

\begin{figure}[!b]
\centering
\includegraphics[scale=0.51]{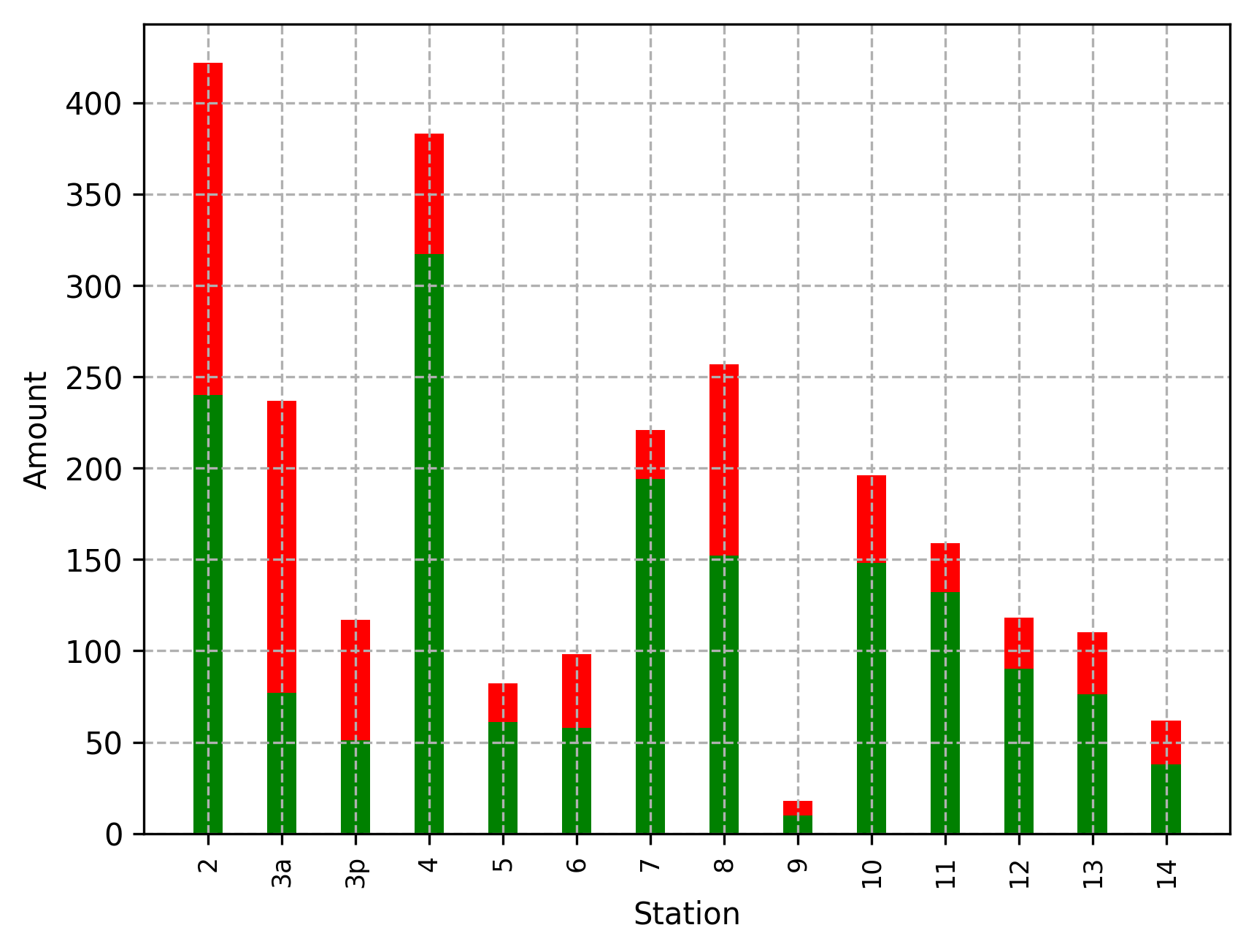}~\includegraphics[scale=0.51]{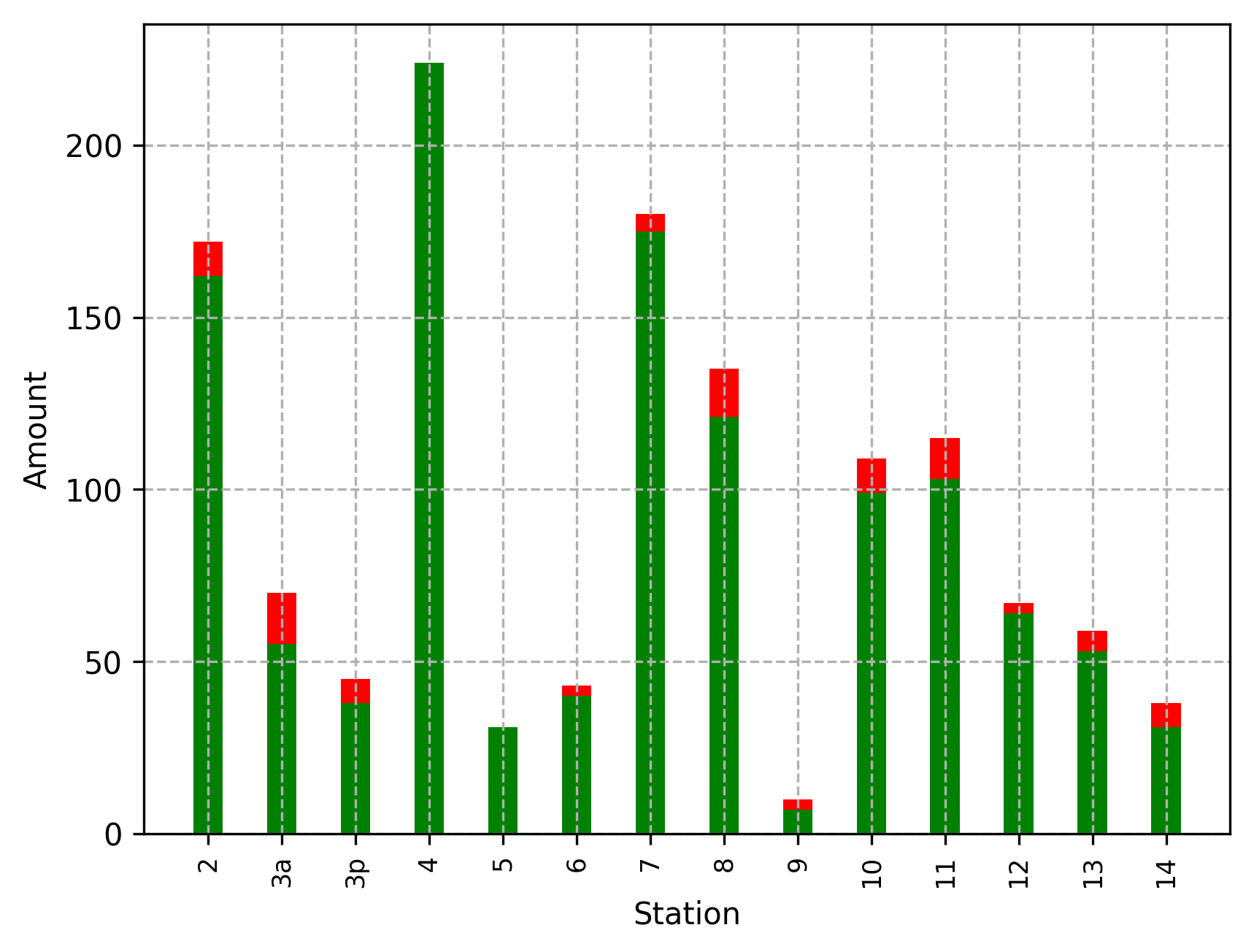}
\caption{Lymph nodes recall performance per primary station for our best approach. To the left all lymph nodes are considered, to the right only the lymph nodes with a short-axis diameter $\geq10$\,mm are considered. The red color represents the total amount of lymph nodes and the green color the amount positively detected.}
\label{fig:recall-per-station}
\end{figure}

From the best performing approach, ensemble of UNet-SW32 and AGUNet-FV-APG, performances based on lymph nodes' primary station and short-axis diameter are reported in Table.~\ref{tab:results-size-stations}.
When considering lymph nodes of all sizes, a $6$\% recall increase can be witnessed when discarding the lymph nodes located in regions not corresponding to any IASLC station or in station 1 (i.e., relevant stations category). Those lymph nodes are either heavily under-represented in our dataset, or simply not included during sample preprocessing, explaining the model's worst performance. A significant recall improvement of $20$\% can be acknowledged when focusing on lymph nodes with a short-axis diameter $\geq7$\,mm. Regarding the most clinically relevant lymph nodes, with a short-axis diameter $\geq10$\,mm and potentially malignant following the RECIST criterion, an overall recall of $90.73$\% is reached along with a patient-wise recall of $92$\%.
The pixel-wise segmentation performances are comparatively worse, with an average Dice-TP of $59$\% at best. However, the score is not directly representative of the segmentation quality and is heavily impacted by the process including the connected components approach and pairing strategy. A Dice score computed between a single lymph node and a cluster of collocated lymph nodes will be inherently low, even if the pixel-segmentation is accurate over the different lymph nodes in the cluster. If anything, this value is an indication of the method's struggles to perform instance disambiguation. Conversely, the GT-Perc score is a better metric to assess the total extent of properly segmented lymph node voxels. On average, up to $80$\% of the voxels belonging to the clinically relevant lymph nodes have been successfully segmented. While the noisiness in the results is not factored in here, the combination of the different metrics (i.e., Dice-TP, GT-Perc, and FPPP) should be enough to appreciate the quality of the pixel-wise segmentation.

Leaving laterality aside, the recall performances are reported with respect to each of the IASLC station in Fig.~\ref{fig:recall-per-station}.
Considering only the clinically relevant lymph nodes (i.e., right sub-plot), no major recall discrepancy can be identified across the different stations. Even though the distribution of lymph nodes featured in each station is heavily unbalanced, especially between station 4 and 9, the model's ability to segment the largest lymph nodes has not been hindered.
When including lymph nodes of all short-axis diameters (i.e., left sub-plot), stations 2, 3, and 8 are standing out with a recall below $60$\%. Station 2 being the most populated station, the lower recall percentage cannot be explained by an under-representation or data imbalance rationalization. Having the possibility to train our models with a higher input resolution should help segmenting the smallest and challenging lymph nodes.

\begin{figure}[!t]
\centering
\includegraphics[scale=0.57]{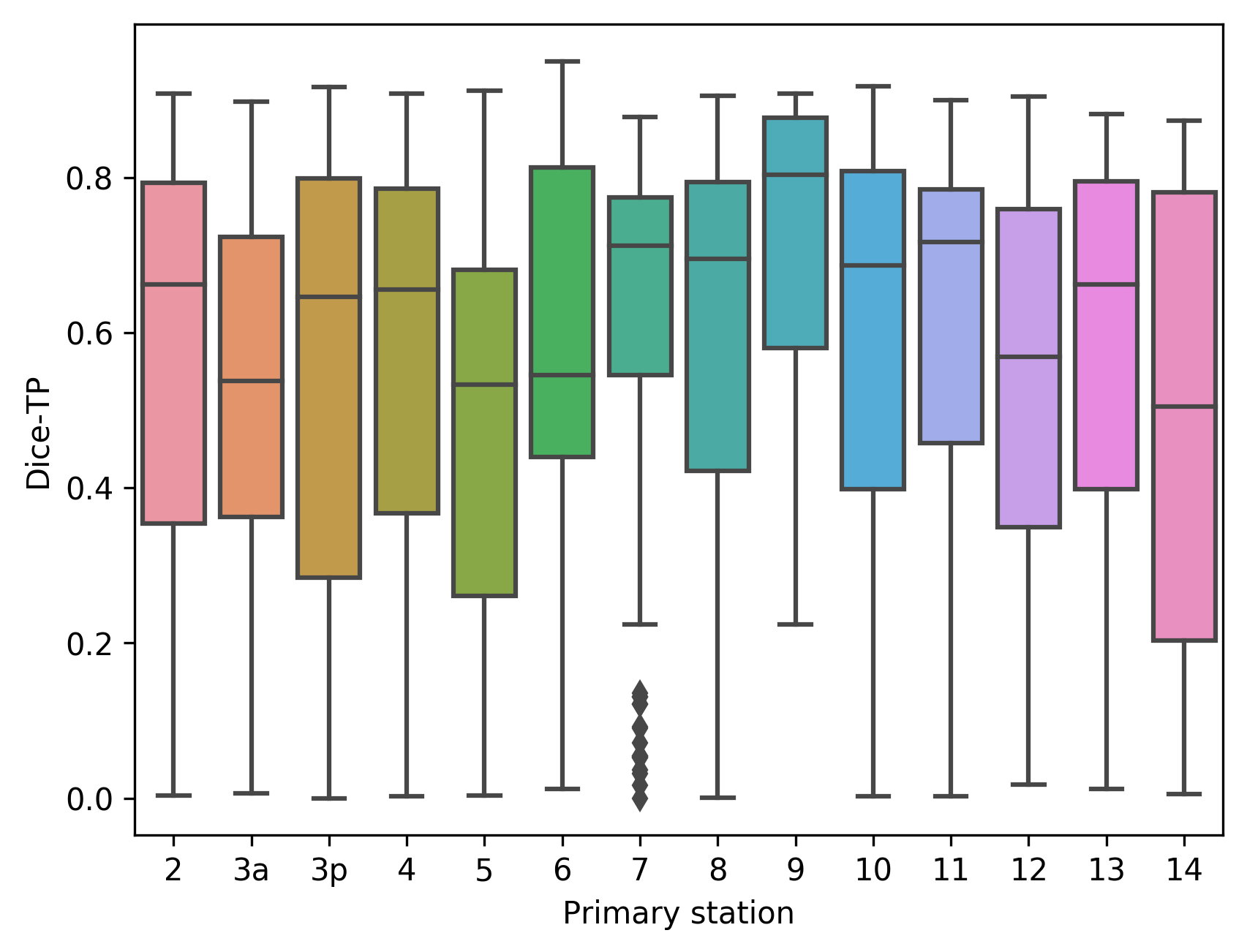}~\includegraphics[scale=0.57]{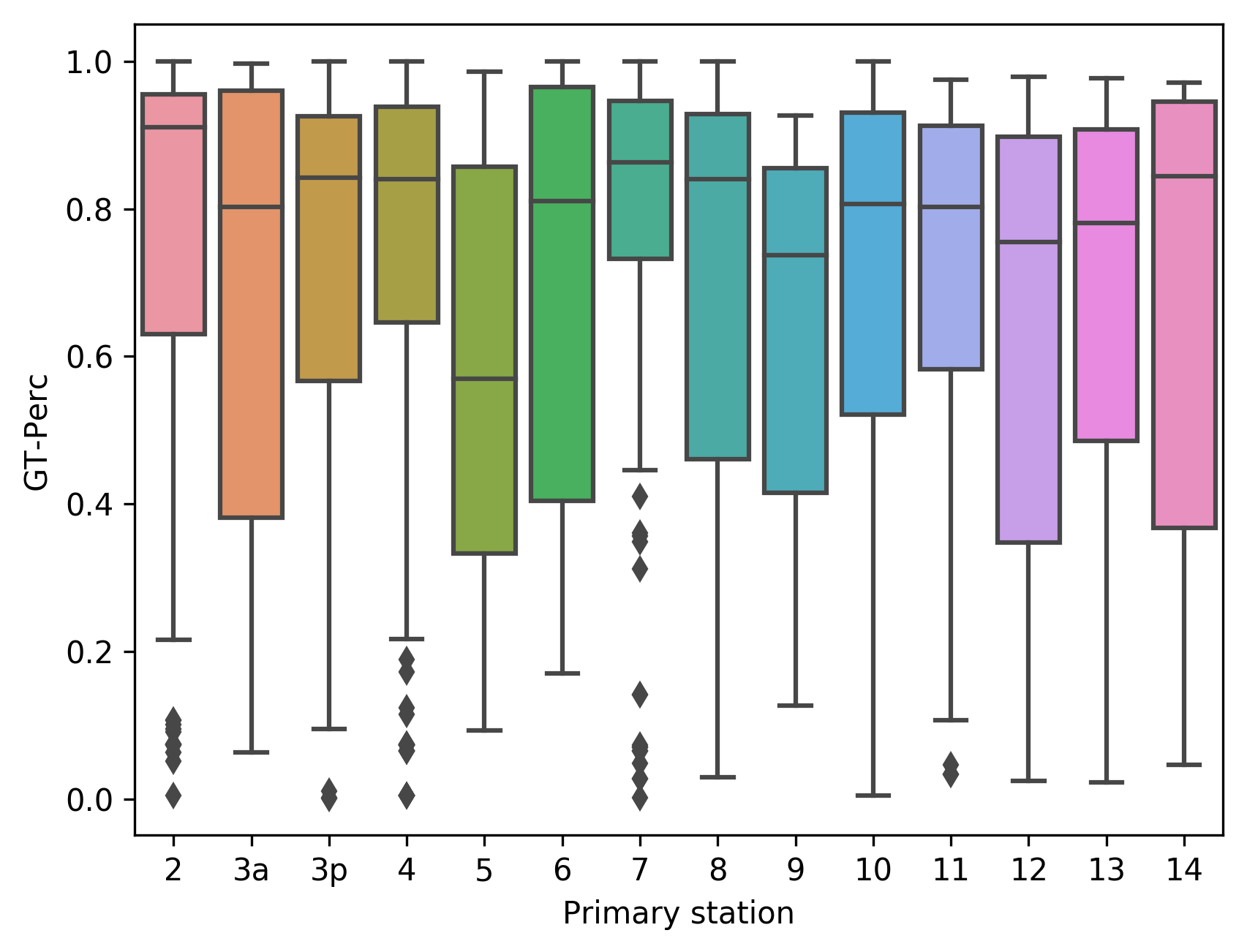}
\caption{Segmentation performance for lymph nodes with a short-axis diameter $\geq10$\,mm and featured in the relevant stations category, obtained with our best performing approach. The Dice-TP scores are reported to the left and the GT-Perc scores to the right.}
\label{fig:detected-dice-per-station}
\end{figure}

An investigation over the pixel-wise segmentation quality across the different stations is provided in Fig.~\ref{fig:detected-dice-per-station}. The average GT-Perc scores (right sub-plot) are stable across the different stations with values around or above $80$\%. Only station 7 seems to be standing out with a more compact range of scores, far from the few outliers. A similar general behaviour can be observed for the Dice-TP scores (left sub-plot).

\subsection{Ground truth quality assessment over a benchmark subset}
For assessing the quality of the ground truth, a direct comparison between assigned stations for the $15$ patients of the benchmark subset has been performed. Out of the $363$ lymph nodes featured,  $312$ instances were attributed a perfect grade as both the expert and medical trainee were in agreement, amounting to about $86$\% of all cases. A good grade was associated with $33$ instances (around $9$\%), as the expert believed the primary and secondary stations assigned by the trainee were to be swapped. Given the loose definition from the IASLC guidelines, capturing exactly the quantity of a lymph node lying in each station is impossible solely from a CT volume. As such, $95$\% of all lymph nodes were considered to be correctly assigned with their main stations (up to three). Lastly, $18$ instances were assigned a wrong station according to the expert, representing slightly less than $5$\% of all cases.
The vast majority of confused cases revolved around a mistaken assignment to station 10 whereby lymph nodes should have been assigned to station 4 (in 7 cases), station 5 (in 4 cases), or station 6 and 12 (in 1 case each).

Regarding the quality of the pixel-wise segmentation, the task was not asked to be performed by the second expert, as being too time-consuming. From a thorough eye-balling, no clinically relevant lymph node was mentioned as having been overlooked. On three occurrences, the segmentation was suggested to be refined as more than one lymph node could be identified from the segmented cluster.

For completeness, the performances obtained on the benchmark subset, with our best performing method, are reported in Tab.~\ref{tab:results-size-stations_benchmark}. Not indicated in the table, an FPPP rate of $4.73\pm2.52$ was obtained on average over the $15$ patients.

\iffalse
\begin{table}[!t]
\caption{Segmentation performances obtained with our best performing method for the three lymph node categories based on short-axis value. The first four measurements (left-most columns) are reported for all lymph nodes while the last four measurements (right-most columns) are reported for the lymph nodes belonging to relevant IASLC stations. Only for the 15 patients of the benchmark subset.}
\adjustbox{max width=\textwidth}{
\begin{tabular}{c||c|c|c|c||c|c|c|c|}
\multirow{2}{*}{Short-axis} & \multicolumn{4}{c||}{All stations} & \multicolumn{4}{c|}{Relevant stations} \tabularnewline
\cline{2-9}
 & Dice-TP & GT-Perc & Recall & Recall-PW & Dice-TP & GT-Perc & Recall & Recall-PW \tabularnewline
\hline
All & $44.75\pm13.52$ & $53.13\pm19.44$ & $46.42$ & $46.47\pm11.98$ & $44.49\pm13.07$ & $52.61\pm19.24$ & $56.02$ & $58.39\pm16.19$ \tabularnewline
$\geq7\,mm$ & $47.92\pm12.89$ & $55.78\pm20.74$ & $72.67$ & $80.01\pm19.58$ & $47.66\pm12.50$ & $55.26\pm20.63$ & $78.26$ & $81.78\pm17.04$ \tabularnewline
$\geq10\,mm$ & $53.42\pm16.70$ & $58.86\pm22.78$ & $82.72$ & $88.60\pm15.57$ & $53.14\pm16.47$ & $58.30\pm22.75$ & $89.19$ & $90.49\pm13.11$ \tabularnewline
\end{tabular}
}
\label{tab:results-size-stations_benchmark}
\end{table}
\fi

\begin{table}[!t]
\caption{Segmentation performances obtained with our best performing method for the three lymph node categories based on short-axis value. The first four measurements (left-most columns) are reported for all lymph nodes while the last four measurements (right-most columns) are reported for the lymph nodes belonging to relevant IASLC stations. Only for the 15 patients of the benchmark subset.}
\adjustbox{max width=\textwidth}{
\begin{tabular}{c|cccc|cccc|}
\multirow{2}{*}{Short-axis} & \multicolumn{4}{c|}{All stations} & \multicolumn{4}{c|}{Relevant stations} \tabularnewline
\cline{2-9}
 & Dice-TP & GT-Perc & Recall & Recall-PW & Dice-TP & GT-Perc & Recall & Recall-PW \tabularnewline
\hline
All & $44.75\pm13.52$ & $53.13\pm19.44$ & $46.42$ & $46.47\pm11.98$ & $44.49\pm13.07$ & $52.61\pm19.24$ & $56.02$ & $58.39\pm16.19$ \tabularnewline
$\geq7\,mm$ & $47.92\pm12.89$ & $55.78\pm20.74$ & $72.67$ & $80.01\pm19.58$ & $47.66\pm12.50$ & $55.26\pm20.63$ & $78.26$ & $81.78\pm17.04$ \tabularnewline
$\geq10\,mm$ & $53.42\pm16.70$ & $58.86\pm22.78$ & $82.72$ & $88.60\pm15.57$ & $53.14\pm16.47$ & $58.30\pm22.75$ & $89.19$ & $90.49\pm13.11$ \tabularnewline
\end{tabular}
}
\label{tab:results-size-stations_benchmark}
\end{table}

\section{Discussion}
\label{sec:discussion}
Solely from the visual inspection of a CT volume, a perfect identification, pixel-wise segmentation, and station mapping of all lymph nodes is arguably close to impossible for various reasons. The resolution in CT acquisitions is relatively correct, but inferior to what ultrasound can achieve during the EBUS procedure. The timing and quality of the contrast uptake is volatile, resulting in some regions visually exhibiting the same characteristics and attenuation as lymph nodes (e.g., mucus, fluid, or other soft tissues), often without clear boundaries. Only a thorough biopsy sample from every lymph node candidate identified on CT would ascertain the perfection of the ground truth, which is unrealistic to achieve in practice. Therefore, a conservative annotation approach was adopted whereby every suspicious region has been labelled as a lymph node, potentially engendering a bias towards hyper-detection. Similarly, visible lymph nodes of any short-axis diameter were annotated if possible, even smaller than the recommended RECIST criterion of $10$\,mm. The inclusion of smaller lymph nodes can be seen as a very efficient data augmentation approach especially for location, either bypassing or complementing heavier data augmentation transforms during training (e.g., zoom, affine, or perspective operations). In addition, clinicians tend to find all PET-positive lymph nodes to be relevant, regardless of size, and often the number and pattern of enlargement in even smaller lymph nodes is looked at.\\
Regarding station mapping, the guidelines prescribed by the IASLC are approximate since construed relatively to the surrounding anatomical structures in the mediastinum, leaving room for interpretation and speculating. In addition to constant evolution of the guidelines through yearly updates, lymph nodes, either expressed as singular entities or within a cluster, can be featured in multiple stations simultaneously.
For all those reasons, we do not claim exact segmentation and station mapping in our ground truth. However, from the results of the fourth validation study, the deviation between the medical trainee and expert was minimal. Some trust regarding the manual annotations can then be granted for the other patients not proofed by the expert. In any case, for this nontrivial task a perfect ground-truth cannot be expected, and we believe this annotation work to be a step in the right direction, sufficient to get insights over the performance in each station. Stemming from this work, multi-task architectures simultaneously performing pixel-wise segmentation, instance detection, and station classification could be investigated.

Regarding the architecture designs or training schemes investigated, a mild impact could be appreciated from the simpler or more complex operations, with a 5-10\% variation across the different metrics studied. The main limitation to get better performances is most likely coming from the dataset itself rather than the chosen methods. While a hefty number of lymph nodes are featured, a total of $120$ patients might not represent enough diversity, especially when training full volume approaches whereby one patient equates to only one training sample. Lymph nodes do exhibit a wide range of expressions (i.e., shape, size, location), a clear imbalance in the stations distributions has been highlighted, and a disparity in contrast-enhancing has also been witnessed. Therefore, access to a dataset orders of magnitude larger would be needed but it represents a tremendous data collection and annotation workload.
Performing ad-hoc ensembling as post-processing enables to benefit from the higher spatial resolution generated by the slab-wise models and global relationships from full volume models. As it is, the ensemble approach is favoring recall and GT-Perc in essence but at the expanse of precision. Smarter and end-to-end consensus designs should be investigated in-depth, using more than simply two models, which would require more data and induce a longer training time.

Using anatomical priors during the training process partly had the expected behaviour, whereby less false positive segmentation was predicted over the esophagus and azygos vein. At the same time, side-effects were also witnessed from lower predicted probabilities over lymph nodes close to those anatomical structures and with similar attenuation values. In order to purely favor instance detection recall, anatomical priors might be leveraged as well in post-processing. The larger number of voxels predicted with a high probability to belong to the lymph node class could be refined by applying a mask containing the location of every other know anatomical structure in the mediastinum. The number of false positives per patient would then be drastically reduced, the pixel-wise segmentation over the lymph nodes refined, and the overall patient-wise recall kept high. With such strategies, well-performing models are required for at least fifteen to twenty anatomical structures. Considering a standalone model for each anatomical structure, the total processing time for a new CT patient would be forcibly longer yet not detrimental as real-time processing is not a requirement for this modality. This post-processing step could either be performed as a simple masking, or end-to-end through a shallow refinement network.
Be it as it may, we believe there is strong potential in further investigating anatomical priors guiding, which would circumvent the need for any refinement or post-processing step. More training samples, and especially the knowledge of more than four other anatomical structures, appear mandatory to proceed.

For the performance assessment, with results generated by a pixel-wise segmentation architecture deprived of any instance detection component, a direct comparison to the raw ground truth was deemed unfair. In many cases involving collocated lymph nodes, no apparent boundary is clearly visible between the different elements of a cluster. The use of a connected component approach over a thresholded prediction map, and the conceivable existence of noise in the pixel-wise segmentation, prevents from a fair one-to-one mapping between prediction and ground truth.
Besides the patient-wise Dice score to assess the quality of the segmentation, reporting instance-wise segmentation performances in two different ways gives further insights. With Dice-TP, the segmentation quality and correctness in instance pairing can be simultaneously assessed. The full extent of segmented ground truth, disregarding instantiating consideration and false positives, can be gauged by GT-Perc scores. Given the challenges in reporting instance detection performances from results of a pure segmentation method, all the metrics considered in the validation studies should be sufficient to cover all aspects and properly enable to put our results in perspective.\\
Given the extremely long time required to perform a full 5-fold cross-validation with slab-wise approaches, nearing 60 days, only a few experiments could reasonably be carried out. Given the low variability in performance across the different folds, comparing architectures and designs over a single fold is arguably enough to identify the best performing ones. Interestingly, models trained using full volumes as input are performing nearly as well as slab-wise approaches, with the benefit to train at a much faster pace. Focusing on improving full volume approaches might be more relevant in the future as experiments can be carried out with less time required.

When considering clinically relevant lymph nodes (i.e., with a short-axis diameter $\geq10$\,mm), and featured in relevant IASLC stations, a patient-wise recall of $92$\%, a segmentation extent of $81$\% on average, and a FPPP ratio of $5$, was reached. Even though the distribution of lymph nodes per station is heavily uneven, the segmentation and detection performances are smooth and stable across all stations. Considering a potential use in a clinical setting, such results are encouraging as a first step to generate report and compute automatic measurements. The main limitation is the inability to properly detect instances and separate collocated lymph nodes. In its current status, our model can be used as preprocessing to bring attention to mediastinal areas potentially featuring lymph nodes, letting the clinical team decide on the mandatory zones where to perform biopsy.

In this work, the knowledge regarding the stations was merely used as a tool to drive the interpretation of the results, with the intent to fine-tune the training schemes in case of need. From the large room given to interpretation, and the proportion of lymph nodes to lie in at least two stations simultaneously, training an end-to-end multi-task architecture for segmentation and station classification appears to this day too challenging. Collecting more data ranks high in the list of future tasks, enabling the possibility to investigate smarter balancing or training sampling schemes. As intermediate solution, station classification could be explored as a refinement step from the results of the pixel-wise segmentation, either in an ad-hoc fashion or by using another shallower architecture. Finally, investigating multi-task architecture in 3D (e.g., Mask R-CNN, YOLACT) is a possibility but would require more powerful GPUs, and a larger dataset.

\section{Conclusion}
\label{sec:conclusion}
The segmentation of mediastinal lymph nodes in CT volumes has been investigated in this study with a focus on 3D neural network architectures and ensembles. Simple and more complex architectures using U-Net as backbone were explored, leveraging either the entire CT volume after heavy downsampling or slab-wise. To benefit from the advantages of each strategy, post-training ensembles enable to generate the best segmentation and instance detection performances. Similarly, anatomical priors guiding reduces false positive predictions over other anatomical structures with similar attenuation values. For clinically relevant lymph nodes, a patient-wise recall of $92$\% was reached for a ratio of up to 5 false positives per patient, with steady performances across the different IASLC stations. In future work, being able to dissociate collocated lymph nodes and properly perform instance detection using a multi-task architecture appears of interest. In addition, better leveraging the station information is of importance to transform the method into a proper and trustworthy diagnostic tool. Finally, increasing the dataset size is needed to gain more data diversity which would in turn improve overall performances.

\subsection*{Disclosures}
The authors declare that the research was conducted in the absence of any commercial or financial relationships that could be construed as a potential conflict of interest.\\
Informed consent was obtained from all individual participants included in the study.

\subsection* {Acknowledgments}
This work was funded by the Norwegian National Advisory Unit for Ultrasound and Image-Guided Therapy (usigt.org) at St. Olavs hospital, Trondheim, Norway.

\bibliographystyle{unsrt}  
\bibliography{references}

\end{document}